# COMPOSING A PUBLICATION LIST FOR INDIVIDUAL RESEARCHER ASSESSMENT BY MERGING INFORMATION FROM DIFFERENT SOURCES

Lucy Amez*

Vrije Universiteit Brussel (VUB), R&D dept., Pleinlaan 2, B-1050 Brussels,
Belgium and Policy Research Centre for R&D Indicators (SOOI)
E-mail: lucy.amez@vub.ac.be,
Phone: +32 (0)2 629.22.21, Fax: +32 (0)2 629.36.40

Nadine Rons

Vrije Universiteit Brussel (VUB), R&D dept., Pleinlaan 2, B-1050 Brussels,
Belgium

* Corresponding author

## 1 Background

Citation and publication profiles are gaining importance for the evaluation of top researchers when it comes to the appropriation of funding for excellence programs or career promotion judgments. Indicators like the Normalized Mean Citation Rate, the h-index or other distinguishing measures are increasingly used to picture the characteristics of individual scholars. Using bibliometric techniques for individual assessment is known to be particularly delicate, as the chance of errors being averaged away becomes smaller whereas a minor incompleteness can have a significant influence on the evaluation outcome. The quality of the data becomes as such crucial to the legitimacy of the methods used.

## 2 Problem

When bibliometric tools are applied for research evaluation, in many cases a publication list is made available to the evaluator. However, it is often presented in a format which is hard to import directly in standard databases. Moreover, the record might include types of publications that will not be taken into consideration, such as contributions in national journals or in e-journals, while at the same time prove incomplete with respect to the categories that do matter. When dealing with expert panel selection, up to date lists of publications are often needed, yet largely unavailable. A reliable publication list then is to be composed which merges information coming from several sources.

Various difficulties are encountered when having to compose a researcher's personal publication record from existing international databases. There is the presence of homonyms, there are name variants for the same author, typing errors or mere incompleteness in data. In cases of identical names, additional variables such as address, affiliation or research field will need to be considered. This is not straightforward as researchers change locations during their careers and sometimes also move to other fields. Recently, clustering algorithms were suggested to separate articles coming from authors holding the same name [WOODING ET AL., 2006]. These can be performed along multiple dimensions like address, subject or co-authorship. Also the Distinct Author Set of Thomson Web of Science uses this type of clustering techniques, combining factors such as cited references, times cited, co-publications and some address information. Besides quantitative methods, more qualitative approaches, like linguistic typology prove valuable in cases where uncertainty remains.

## 3 Methodology

Using a selected group of top researchers, the differences are investigated between publication records composed using 1. WoS distinct author set and 2. Entering key address words traced from the author's CV. The latter aims to maximally capture the different places he/she has worked. In order to verify if other missing articles can be found, both records are compared to freely available publication lists issued by the author, his/her institution or funding agencies.

Even after merging the information coming from different sources, there remains a set of publications, characterised by a considerable degree of uncertainty whether to allocate them to a particular author or not, certainly when they were retrieved from one database but not present in another. In cases where quantitative methods fail to allocate a publication, the study of language turns out a useful method to diminish uncertainty. Syntax form analysis of titles or abstracts appears a rich method when searching for similarities. The style of formulating, the way sentences are build and the phrasing can in some particular cases deliver more certainty than do distance measures when it comes to allocate publications to a specific author.

## 4 Results

The presented work suggests a method to compose a trustworthy publication database using different search techniques. The results provide descriptive statistics with respect to the differences obtained by several approaches and clarifies the main reasons why some articles were detected using one method and not by another. Finally, by means of examples, the paper shows how to link an article to an author based on linguistic typology.

## 5 Conclusion

The paper investigates to what extent publications could be traced by using the Distinct Author Set from WoS compared to a search based on an individual trajectory address list and explains the main reasons causing differences in outcome. Besides that, examples are offered how linguistic analysis can help to diminish uncertainty for linking a particular publication to a single author. Finally, the findings allow to argument and illustrate why for some authors it will always remain difficult to adjust and compose a complete publication list.